\documentstyle[11pt,newpasp,twoside,epsf]{article}
\begin{document}

\title{Modeling the Tidal Tails of the Sagittarius Dwarf Galaxy} 

\author{David R. Law, Steven R. Majewski, Michael F. Skrutskie}
\affil{Dept. of Astronomy, University of Virginia,
Charlottesville, VA 22903-0818}

\author{Kathryn V. Johnston}
\affil{Wesleyan University, Department of Astronomy, Middletown, CT}

\begin{abstract}
N-body simulations are used to model the tidal disruption of the
Sagittarius (Sgr) dwarf galaxy with constraints
set by the positions and velocities of M giants in the Sgr tidal arms 
recently revealed by the Two Micron All-Sky Survey (2MASS).
The simulated Sgr dwarf is placed on a variety of orbits within a
Milky Way potential parameterized by variable circular velocities,
halo flattenings and radial profiles.  
Two hundred separate test particle orbits have been used to explore a wide range of
model Milky Way potentials and dwarf galaxy characteristics.
The family of models is delimited by the data to a relatively narrow allowed range of parameters,
and then input into N-body simulations.
We present our best-fitting model, and discuss the orbital period, apoGalacticon distance, current
space velocity, mass-to-light ratio, and other characteristics of the Sgr dwarf.  In addition, we discuss
the implications of this model for the flattening of the Galactic halo.
\end{abstract}

\section{Introduction}
Since the discovery of the Sgr dwarf by Ibata et al. (1994) many groups
(e.g., Johnston, Hernquist, \& Bolte 1996, Ibata et al. 1997, Ibata \& Lewis 1998, G{\' o}mez-Flechoso, Fux, \& Martinet 1999,
Johnston et al. 1999, Helmi \& White 2001)
have sought to model the Sgr - Milky Way
interaction with respect to a modest patchwork of observational constraints.
Recently, Majewski et al. (2003a, hereafter ``Paper I'') have shown that the extensive length of the Sgr tidal
tails can be traced by M giant stars visible in the all-sky view of the system provided by the 2MASS database.
Spectroscopy of Sgr candidate stars has allowed determination of radial velocities throughout the trailing
tail (Majewski et al. 2003b, hereafter ``Paper II''), and
these substantial new constraints can be used to develop more refined models of the Sgr system.

In this contribution, we briefly describe some of the major results of such modeling.  A comprehensive description
of this new Sgr disruption model can be found in Law, Johnston, \& Majewski (2003, hereafter ``Paper III'').

\section{Modeling the Sgr System}
Following previous work by Johnston et al. (1996, 1999)
the Milky Way potential is represented numerically by a
Miyamoto-Nagai (1975) disk, Hernquist spheroid, and a logarithmic halo.
The total mass and radial profile are fixed by requiring that the rotation curve of this
model Galaxy be consistent with HI \& CO tangent point observations
(e.g., Honma \& Sofue 1997).

The Sgr dwarf itself is represented by $10^5$ self-gravitating particles (representing both the dark and 
light matter components of the satellite),
which are initially distributed according to a Plummer (1911) model.  This satellite
is evolved through the simulated Galactic potential for five orbital periods
using a self-consistent field code (Hernquist \& Ostriker 1992).
The present-day simulated dwarf is constrained to be located at 
$(l,b) = (5.6^{\circ},-14.2^{\circ})$ at a solar distance of $D_{\rm Sgr} = 24$ kpc (Paper I, Ibata et al. 1995)
and have a radial velocity of $v_{\rm LOS,Sgr} = 171$ km s$^{-1}$ (Ibata et al. 1997).
The direction of the dwarf's space velocity vector is determined by requiring that the dwarf orbit in the orbital
plane observed in Paper I.

Subject to these requirements, test-particle orbits 
(i.e. orbits calculated for a test particle with the observed kinematical characteristics of Sgr) and N-body simulations
are performed for simulated satellites with a variety of orbital speeds.  These simulations can be additionally constrained
using the 2MASS M giant distance and radial velocity data presented in Papers I and II. 
Fig. 1 compares the M giant data (Panels a-b, filled squares) with the model Sgr dwarf whose 
tidal tails best reproduce the observations (Panels c-d).  Note the close agreement between model
and observed debris distances and radial velocities along the trailing debris tail 
($\Lambda_{\odot} = 0^{\circ}$ - $100^{\circ}$)\footnote{We use the orbital longitude coordinate
system in the Sgr orbital plane defined in Paper I.}.  This best-fit model
is characterized by a period of 0.75 Gyr with apoGalacticon 52 kpc, 
periGalacticon 14 kpc, and a present space velocity of $(U,V,W) = (237.2, -43.4, 218.9)$ km s$^{-1}$.

\begin{figure}
\plotone{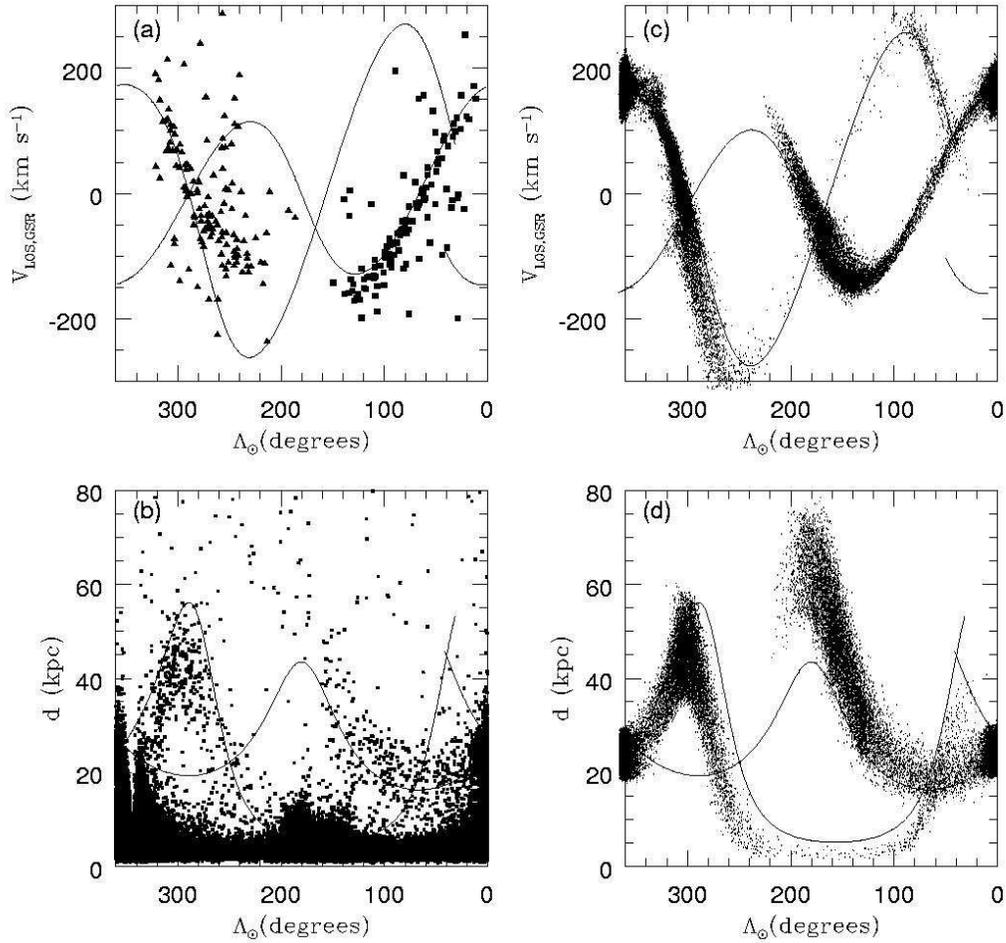}
\caption{Distance and radial velocity are plotted as function of orbital longitude for
M giant data from Papers I \& II (Panels (a) - (b), filled squares) and model Sgr debris unbound from the 
satellite over the last three pericentric passages (Panels (c) - (d)).
The solid line in all panels represents the orbit of the model Sgr dwarf core.  Filled triangles in panel (a) represent
new radial velocity data presented in Paper IV.}
\end{figure}
Although we do not attempt to model the Sgr core in detail, it is nonetheless possible to use the width of the Sgr debris
stream to estimate such global characteristics as the bound mass of the dwarf.
The simulated dwarf which appears to best fit the width of streams shown in Fig. 1 has a present mass of
$M_{\rm Sgr} = 3 \times 10^8 M_{\odot}$ and a mass-to-light ratio $M_{\rm Sgr}/L_{\rm Sgr} = 21$.

\section{Discussion}
As demonstrated in the previous section, the tidal tails of this model provide a good fit to the all-sky view of M giants
presented in Papers I and II.  It is therefore possible to use this model to determine what range of Milky Way models
permit simulated satellites to reproduce observations.  Particularly, N-body simulations can be used to constrain the
flattening of the Galactic halo (e.g., Ibata et al. 2001).  
Fitting an orbital plane to leading and trailing M giant debris separately, we determine
that the orbital pole of Sgr debris has precessed by $1.7^{\circ} \pm 2.4^{\circ}$ over about $300^{\circ}$ of orbital
longitude.  Repeating this calculation for N-body simulations in model dark halos with a variety of flattenings, we calculate
pole precessions of  $2.2^{\circ} \pm 1.6^{\circ}$,
$3.5^{\circ} \pm 1.7^{\circ}$, and $5.6^{\circ} \pm 1.4^{\circ}$ for flattenings in the halo potential of
$q = 1, 0.95$, and $0.90$ respectively.  It therefore appears likely that the halo of the Milky Way 
can be described by an almost spherical potential.

Although this model provides a good match to the distances and velocities of trailing Sgr debris given in Papers I and II,
it does not fit recent data obtained by Majewski et al. (2003c, hereafter ``Paper IV'') in the region of the 
Sgr leading arm.  Fig. 1 (Panel a, filled triangles) plots these new data, which has velocities slower than that of the model
by up to 200 km s$^{-1}$ in the range $\Lambda_{\odot} = 300^{\circ}$ - $200^{\circ}$.  There is no simple modification of the
velocity of the model satellite that serves to reproduce this new trend, and this may be an indication of such
other effects as dynamical friction.  However, simulations suggest that including corrections from Chandrasekhar's
formulation of dynamical friction
should not have a substantial effect on the observed velocities of leading tidal debris for model satellites with
mass $M_{\rm Sgr, 0} \leq 10^{10} M_{\odot}$, and we find that accurately reproducing the observed trend
is difficult even for satellites with initial masses greater than this.

This inconsistency and implications of the best-fit model for the size and shape of the Milky Way are
discussed at greater length in Paper III.

\acknowledgements
DRL acknowledges support from the Local Organizing Committee, a U.Va. Small Research Fellowship, and the U.Va. Echols Program.
The authors also acknowledge NASA/JPL contracts 1228235 (SRM/KVJ) and 1234021 (MFS).

\end{document}